\documentclass[preprint2,twoside]{hwo}

\usepackage{graphicx}

\input{hwo.h}

\begin{document}

\title{\textbf{\LARGE Probing the Origin of Water in Planets within Habitable Zones by HWO}}
\author {\textbf{\large Yasuhiro Hasegawa,$^{1}$ Courtney Dressing,$^2$ and Ludmila Carone$^3$}}
\affil{$^1$\small\it Jet Propulsion Laboratory, California Institute of Technology, Pasadena, CA 91109, USA; \email{yasuhiro.hasegawa@jpl.nasa.gov}}
\affil{$^2$\small\it Department of Astronomy, University of California, Berkeley 501 Campbell Hall \#3411Berkeley, CA 94720, USA.}
\affil{$^3$\small\it Space Research Institute, Austrian Academy of Sciences, Schmiedlstrasse 6, 8042, Graz, Austria.}





\begin{abstract}
How do habitable environments arise and evolve within the context of their planetary systems? 
This is one fundamental question, and it can be addressed partly by identifying how planets in habitable zones obtain water. 
Historically, astronomers considered that water was delivered to the Earth via dynamical shake-up by Jupiter, 
which took place during the formation and post-formation eras (e.g., $\lesssim 100$ Myr). 
This hypothesis has recently been challenged by a more dynamic view of planet formation; 
planet-forming materials move in protoplanetary disks via various physical processes such as pebble drift and planetary migration. 
\textit{Habitable Worlds Observatory} (HWO) will open a new window to address this important, but difficult question 
by discovering and characterizing Earth-like exoplanets around G-type stars.
In this article, we consider two possible working hypotheses:
(1) the abundance of water on planets in habitable zones has \textit{any} correlation with the presence of outer planets; 
and (2) the abundance of water on planets in habitable zones has \textit{no} correlation with the presence of outer planets.
We discuss what physical parameters need to be measured to differentiate these two hypotheses and 
what observational capabilities are desired for HWO to reliably constrain these physical parameters.
\\
\\
\end{abstract}

\vspace{2cm}

\section{Science Goal}

Identifying how habitable environments emerge and evolve is one key information to quantify the ubiquity of habitable planets in the Universe. 
The science case described below that may be able to be explored by HWO will therefore enable astronomers to scientifically address one of the fundamental questions of humanity: 
”Are we alone?” 

\subsection{Topics Related to the Astro2020:}

This science case is relevant to the following Key Science Questions and Discovery Areas of the Astro2020 Decadal Survey Report:

\begin{itemize}
    \item \textbf{E-Q3.} How do habitable environments arise and evolve within the context of their planetary systems? 
    \begin{itemize}
        \item \textbf{E-Q3a.} How are potentially habitable environments formed?
        \item \textbf{E-Q3b.} What processes influence the habitability of environments?
        \item \textbf{E-Q3d.} What are the key observable characteristics of habitable planets?
    \end{itemize}

 \end{itemize}

\section{Science Objective}

The origin of water on the Earth remains elusive.
Historically, astronomers considered that water on the Earth was likely delivered from the asteroid and Kuiper belt regions \citep[e.g.,][]{2000M&PS...35.1309M}.
This was motivated by the fact that the Earth is extremely water-poor (i.e., $0.07-0.23$ wt\%).

The key underlying assumptions of this hypothesis are 1) that the location of the water snow line was around $2-3$ au at the era of planet formation 
and 2) that the position of Jupiter was at the current position (i.e., 5.2 au).
Water could be delivered to the Earth from the formation era to the post-formation era as the form of embryos/planetesimals from the asteroid belt regions as well as 
the form of planetesimals from the Kuiper belt region.

There are a number of unknowns to tightly constrain the origin of water on the Earth. Below are some examples:
\begin{itemize}
    \item The composition of planet-forming materials (i.e., planetesimals and embryos) remains poorly constrained.
             While astronomical observations and sample return missions (e.g., Hayabusa, OSIRIS-REx) provide useful constraints 
             \citep[e.g.,][]{2011ARA&A..49..471M,2011Sci...333.1116Y,2019Natur.568...55L},
             these constraints are not sufficient to develop a detailed statistical understanding of the properties of planet-forming materials.
    \item  The position of the snow line could move during the formation era, and the position (and eccentricity) of Jupiter could evolve with time as well.
              The discovery of exoplanets challenges the canonical theory of planet formation \citep[e.g.,][]{1995Natur.378..355M,2007ARA&A..45..397U,2010Sci...327..977B}.
              The recent dynamic view of planet formation suggests 
              the movement of planet-forming materials via inward pebble drift and  planetary migration during the formation era and 
              poses serious questions on the two fundamental assumptions adopted in previous studies.
    \item The lack of detections of Earth-like exoplanets around G-type stars hinders testing theoretical predictions.
             While detailed simulations could be run \citep[e.g.,][]{2004Icar..168....1R}, 
             the predictive power of such simulations cannot be quantified 
             partly because the value of model parameters to be used in simulations is poorly constrained as describe above,
             and partly because no observations (except for the solar system) are available for testing the outcome of simulations. 
\end{itemize}

\subsection{Primary Science Objective}

The origin of water on planets in habitable zone is one important, but difficult question to address. 
However, HWO will open a new window.
The main objective of this science case is therefore attempt to test two possible working hypotheses: \\

{\bf The abundance of water on planets in habitable zones has \textit{any} correlation with the presence of outer planets}; or \\

{\bf The abundance of water on planets in habitable zones has \textit{no} correlation with the presence of outer planets}.

\section{Physical Parameters}

A number of physical parameters are involved to test the above hypotheses. Below is the list of these parameters.

\subsubsection*{The number of planets to be detected}

To conduct hypothesis testing, an enough number of exoplanets both in habitable zones and beyond need to be detected.

\begin{itemize}
    \item Under the assumption that 10\% of candidate planets being observed that are actually Earth-like, 
             {\bf about 30 candidates} need to be observed for the robust detection of a single Earth-like planet \citep{2019luvo.rept......}. 
             Given that the current occurrence rate estimate of eta-Earths is about 0.3 \citep[e.g.,][]{2018ApJ...856..122K,2022AJ....164..190B}, 
             {\bf at least 100 targets} need to be observed.
    \item The current occurrence rate estimate of distant massive planets by direct imaging is about a few \% around 5 au \citep[e.g.,][]{2019AJ....158...13N}; 
             if 100 targets will be observed, then {\bf about few distant planets} will be discovered.
\end{itemize}

\subsubsection*{Planet properties}

It is very challenging to reliably estimate the abundance of water in planets. 

\begin{itemize}
    \item One possible approach may be to infer the water abundance from {\bf planet mass and radius}. 
             This approach may work if the water abundance is high enough that the bulk composition of corresponding planets deviates from the Earth-like composition 
             in the mass-radius diagram.
    \item If the water content is very small like the Earth, the above approach won’t work, and another approach is needed. 
             One tentative approach may be to detect {\bf water vapor in exoplanet atmospheres.}
\end{itemize}

If the water abundance will be inferred from planet mass and radius, 
then accurate determination of these two quantities is necessary. 
For instance, if planet mass is measured at infinite accuracy (which is not possible) and 
if the water mass fraction needs to be inferred within the 10 \% accuracy, then the radius measurement tolerates only
less than the 5\% relative error, using the mass-radius relation suggested by \citet{2019PNAS..116.9723Z}.

\subsubsection*{The position of distant planets is another key quantity}

The conventional disk model infers the position of the water snow line to be around $2-3$ au around G-type star \citep[e.g.,][]{1981PThPS..70...35H}.

\begin{itemize}
    \item For this case, water delivery by dynamical shake-up would be most efficient when distant planets are located around {\bf 5 au} due to gravitational resonant interaction
             \citep[e.g.,][]{2000M&PS...35.1309M,2004Icar..168....1R}.
    \item On the contrary, exoplanet observations infer that the orbital architecture of planetary systems could have huge diversity. 
             It might thus be useful to consider {\bf two semimajor-axis bins} (5-10 au and beyond) to quantify any correlations.
\end{itemize}

\section{Description of Observational Requirements}

Two kinds of observations will be needed to achieve our science case.

The first kind of observations are {\bf imaging} of planets at visible with {\bf coronagraph} that will achieve:
\begin{itemize}
    \item the contrast ratio of $\sim 4 \times 10^{-11}$,
    \item the inner working angle (IWA) of $< 80$ mas at 1 $\mu$m, i.e., $> 2 \lambda/D$ or $ \sim 0.8$ au at 10 pc,
    \item the outer working angle (OWA) of $> 1$ arcsec at 0.5 $\mu$m, i.e., $< 80 \lambda/D$ or $ \sim 10$ au at 10 pc, and
    \item the astrometric accuracy of $ \sim 1$ cm s$^{-1}$, i.e., $ \sim0.3 ~ \mu$arcsec,
\end{itemize}
where $D$ is the mirror diameter.
The last requirement will allow measuring the planet mass of $\sim 1 M_{\oplus}$ \citep{2020arXiv200106683G}.

Future ELT observations would be complementary to discover and characterize distant planets. 
$N-$body simulations suggest that observations of planets at about {\bf eight different epochs} would be needed to reliably constrain orbital parameters 
(e.g., semimajor-axis, eccentricity, inclination) of planets \citep[e.g.,][]{2020arXiv200106683G,2025arXiv250721443S}. 

The second kind are {\bf spectroscopic observations} of planets in habitable zones that will achieve:
\begin{itemize}
    \item the wavelength coverage of $0.7 - 1.5 ~\mu$m, and
    \item the spectral resolution of $R > 40$.
\end{itemize}    
These observations will detect {\bf at least two water lines} 
because atmospheric water vapor has five broad spectral absorption features within the wavelength range \citep{2020arXiv200106683G}.

{\bf Acknowledgements.} 
This research was carried out at the Jet Propulsion Laboratory, California Institute of Technology, 
under a contract with the National Aeronautics and Space Administration (80NM0018D0004). 
Y.H. is supported by JPL/Caltech.

\bibliography{author.bib}

\end{document}